\documentclass[11pt,a4paper]{article}

\usepackage{epsfig,amsmath,amssymb,cite}

\begin{document}

\title{Shadow of black holes with a plasma environment \\ in 4D Einstein-Gauss-Bonnet gravity}

\author{Javier Bad\'ia$^{1, 2}$\thanks{e-mail: jbadia@iafe.uba.ar} and Ernesto F. Eiroa$^{1}$\thanks{e-mail: eiroa@iafe.uba.ar}\\
	{\small $^1$ Instituto de Astronom\'{\i}a y F\'{\i}sica del Espacio (IAFE, CONICET-UBA),}\\
	{\small Casilla de Correo 67, Sucursal 28, 1428, Buenos Aires, Argentina}\\
	{\small $^2$ Departamento de F\'{\i}sica, Facultad de Ciencias Exactas y Naturales,} \\ 
	{\small Universidad de Buenos Aires, Ciudad Universitaria Pab. I, 1428, Buenos Aires, Argentina}}
\date{}

\maketitle

\begin{abstract}
We study the shadow cast by rotating black holes surrounded by plasma in the context of the 4D Einstein-Gauss-Bonnet theory of gravity. The metric for these black holes results from applying the Newman-Janis algorithm to a spherically symmetric solution. We obtain the contour of the shadow for a plasma frequency model that allows a separable Hamilton-Jacobi equation. We introduce three observables in order to characterize the position, size, and shape of the shadow.
\end{abstract}


\section{Introduction}

It is well known that Einstein-Gauss-Bonnet theory in four spacetime dimensions is purely topological and thus equivalent to general relativity. A novel gravity theory was recently proposed \cite{glavan20} by rescaling the coupling constant and taking the limit $D \to 4$ in the $D$-dimensional Einstein-Gauss-Bonnet theory, with the intention of overcoming this standard result. But this approach, based on some particular solutions, lacks of a complete set of field equations and does not have an intrinsically four-dimensional description in terms of a covariantly conserved rank-2 tensor in four dimensions \cite{gst20}. A theory with a Gauss-Bonnet term in four dimensions that provide a solution to these problems was then presented \cite{fernandes20,hennigar20}, which propagates a scalar field in addition to the metric tensor and the full action belongs to the Horndeski class of scalar-tensor theories of gravity. The action, obtained by a regularization procedure in a way that is free from divergences, produces well behaved second-order field equations. This theory, usually dubbed regularized or scalar-tensor 4D Einstein-Gauss-Bonnet gravity (4DEGB), includes a nonvanishing contribution coming from the Gauss-Bonnet term \cite{fernandes20,hennigar20}.

The number of theoretical studies on black hole shadows \cite{bardeen73,falcke00} has sharply increased since the Event Horizon Telescope (EHT) \cite{eht19L1,eht19L5} collaboration obtained the first reconstructed image of the supermassive black hole M87*, located at the center of the giant elliptical galaxy M87. Many articles have been published on this topic in recent years, here we can mention only a few of them \cite{tsukamoto18,badia20,lima20,perlick21}, in which more references can be found. The presence of plasma surrounding a black hole affects the characteristics of the shadow \cite{perlick17,yan19,badia21}, because photons with different frequencies follow distinct trajectories, resulting in chromatic effects. In this work, we consider the shadow due to rotating black holes \cite{kumar20,wei21} in the scalar-tensor 4DEGB gravity, within a plasma environment. We define three observables to describe how the presence of plasma affects the size, the deformation and the position of the shadow. We analyze in detail the case of a Shapiro plasma, with the shadow seen by an equatorial observer. We use units such that $G=c=\hbar=1$.

\section{Black holes in 4D Einstein-Gauss-Bonnet gravity}

The scalar-tensor 4DEGB action reads \cite{fernandes20,hennigar20}
\begin{equation} 
    S=\int_{\mathcal{M}} d^4 x \sqrt{-g} \Big[ R + \gamma \Big(4G^{\mu \nu}\nabla_\mu \phi \nabla_\nu \phi - \phi \mathcal{G} + 4\Box \phi (\nabla \phi)^2 + 2(\nabla \phi)^4\Big) \Big] + S_m  .
\label{action}
\end{equation}
The theory is free of divergences and belongs to the Horndeski class, with functions $G_2=8 \gamma X^2$, $G_3=8 \gamma X$, $G_4=1+4 \gamma X$, and $G_5 = 4 \gamma \ln X$, where $X=-(1/2) \nabla_{\mu} \phi \nabla^{\mu} \phi$. The field equations are obtained \cite{fernandes20,hennigar20} by varying this action with respect to the metric
\begin{equation}
    G_{\mu \nu} =  \gamma \mathcal{H}_{\mu \nu} +T_{\mu \nu} , 
\end{equation}
where $\mathcal{H}$ is a complicated second order function in the derivatives of the scalar field $\phi$; there is another equation that is found \cite{fernandes20,hennigar20} by varying with respect to the scalar field $\phi$. The expressions of these equations are not relevant for our purposes.

The 4DEGB field equations in vacuum admit the spherically symmetric and asymptotically flat solution \cite{glavan20,hennigar20} with the metric given by 
\begin{equation}
	ds^2 = - \left(1 - \frac{2m(r)}{r}\right) dt^2 +  \left(1 - \frac{2m(r)}{r}\right)^{-1} dr^2 + r^2 (d\theta^2 + \sin^2\theta\, d\varphi^2),
\label{metric-sphe}
\end{equation}
where 
\begin{equation} 
	m(r) = \frac{r^3}{64\pi \gamma} \left(\sqrt{1 + \frac{128\pi \gamma M}{r^3}} - 1\right).
\label{m-function}
\end{equation}
The parameter $\gamma$ of the theory has units of mass squared. We only consider $\gamma >0$, since the square root becomes imaginary for a finite value of $r$ if $\gamma <0$. The limit $\gamma \to 0$ corresponds to the Schwarzschild geometry. The black hole mass is $M$, since we have $m(r) \to M$ as $r \to \infty$.

A rotating solution was subsequently found by applying a modified version of the Newman-Janis algorithm \cite{kumar20,wei21}, having in the Boyer-Lindquist coordinates the form
\begin{equation}
ds^2 = - \frac{\rho^2\Delta}{\Sigma} dt^2 + \frac{\Sigma \sin^2\theta}{\rho^2} \left[d\varphi - \frac{2a m(r) r}{\Sigma} dt\right]^2 + \frac{\rho^2}{\Delta} dr^2  + \rho^2 d\theta^2,
\label{metric-rot}
\end{equation}
where  $a=J/M$ is the rotation parameter and
\begin{eqnarray}
	\rho^2 &=& r^2 + a^2 \cos^2\theta, \nonumber \\
	\Delta &=& r^2 - 2m(r)r + a^2, \nonumber \\
	\Sigma &=& (r^2+a^2)^2 - a^2 \Delta \sin^2\theta.
\end{eqnarray}
The limit $\gamma \to 0$ recovers the Kerr black hole. The radius $r_{h}$ of the event horizon corresponds to the largest real and positive solution of the equation $\Delta =0$. It can be seen numerically that the condition $\gamma/M^2 < 0.00129$ is necessary for the existence of the event horizon, otherwise there is a naked singularity.

\section{Photon geodesics in a plasma environment}

The motion of photons in a pressureless, nonmagnetized plasma is governed by the Hamiltonian \cite{synge60}
\begin{equation}
	\mathcal{H} = \frac{1}{2} \left( g^{\mu\nu}(x) p_\mu p_\nu + \omega_p(x)^2 \right),
\end{equation}
with $g^{\mu\nu}$ the inverse of the metric tensor, $x^\mu$ the spacetime coordinates, $p^\mu$ the conjugate momenta, and $\omega_p$ the plasma electron frequency
\begin{equation}
	\omega_p^2 = \frac{4\pi e^2}{m_e} N,
\end{equation}
where $e$ and $m_e$ are the electron charge and mass respectively, and $N$ the electron number density. We assume that the plasma frequency is stationary and axisymmetric, so it does not depend on the Boyer-Lindquist coordinates $t$ and $\varphi$. As it happens for the Kerr geometry \cite{perlick17}, the inequality $\omega_0^2 \geq - g_{tt} \omega_p^2$ is a necessary and sufficient condition for a light ray with frequency $\omega_0$ (measured by an observer at infinity) to exist at a given point of the spacetime. 

\subsection{Hamilton-Jacobi equation}

We can write down the Hamilton-Jacobi (HJ) equation for photons, which reads
\begin{equation}
	\mathcal{H}\left(x, \frac{\partial S}{\partial x}\right) = 0
\end{equation}
and we introduce the ansatz
\begin{equation}
	S = - \omega_0 t + p_\varphi \varphi + S_r(r) + S_\theta(\theta).
\end{equation}
By substituting it in the HJ equation, we arrive at
\begin{gather}
	 \Delta (S_r')^2 - \frac{1}{\Delta} \left[(r^2+a^2)^2 \omega_0^2 + 4am(r)r \omega_0 p_\varphi + a^2 p_\varphi^2\right] \nonumber \\
	 + (S_\theta')^2 + a^2 \omega_0^2 \sin^2\theta + \frac{p_\varphi^2}{\sin^2\theta} + \rho^2 \omega_p^2 = 0,
\end{gather}
where the prime denotes the derivative with respect to the corresponding coordinate. The HJ equation is separable if and only if the plasma frequency takes the form \cite{perlick17}
\begin{equation}
	\omega_p^2 = \frac{f_r(r) + f_\theta(\theta)}{\rho^2},
\end{equation}
with $f_r(r)$ and $f_\theta(\theta)$ arbitrary functions of the coordinates $r$ and $\theta $, respectively. The spacetime is axisymmetric, stationary, and asymptotically flat, therefore the quantities $p_t = -\omega_0$ and $p_\varphi$ are conserved along the geodesics of photons; $E=\omega_0$ is the photon energy and $p_\varphi$ is the $z$ component of the angular momentum. Since $\omega_p$ only depends on the coordinates $r$ and $\theta$, the quantities $E$ and $p_\varphi$ are still conserved in the presence of plasma. A third constant of motion for photons is $\mathcal{H} =0$. By substituting  $\omega_p$ into the Hamilton-Jacobi equation, we can separate it as
\begin{equation}
	(S_\theta')^2 + \left(a \omega_0 \sin\theta - \frac{p_\varphi}{\sin\theta}\right)^2 + f_\theta = -\Delta (S_r')^2 + \frac{1}{\Delta} \left[\omega_0(r^2+a^2) - a p_\varphi\right]^2 - f_r.
\end{equation}
Since the left hand side is a function only of $\theta$ and the right hand side a function only of $r$, they should both be equal to a constant $\mathcal{K}$. For convenience, we use instead the Carter constant \cite{carter68}, defined by $\mathcal{Q} = \mathcal{K} - (p_\varphi - a\omega_0)^2$, as the fourth constant of motion. From the Hamilton equations, using that $\dot{x}^\mu = p^\mu = g^{\mu\nu}p_\nu$ (the dot denotes the derivative with respect to the curve parameter $\lambda$) and $ p_ {\nu} = \partial S / \partial x ^ {\nu}$, the equations of motion read
\begin{eqnarray}
	\rho^2 \dot{t} &=& \frac{r^2 + a^2}{\Delta}P(r) - a(a\omega_0\sin^2\theta - p_\varphi), \\
	\rho^2 \dot{\varphi} &=& \frac{a}{\Delta}P(r) - a\omega_0 + \frac{p_\varphi}{\sin^2\theta}, \\
	\rho^2 \dot{r} &=& \pm \sqrt{R(r)}, \\
	\rho^2 \dot{\theta} &=& \pm \sqrt{\Theta(\theta)},
\end{eqnarray}
where
\begin{eqnarray}
	R(r) &=& P(r)^2 - \Delta [\mathcal{Q} + (p_\varphi - a \omega_0)^2 + f_r], \\
	\Theta(\theta) &=& \mathcal{Q} + \cos^2\theta \left(a^2 \omega_0^2 - \frac{p_\varphi^2}{\sin^2\theta}\right) - f_\theta,  \\
	P(r) &=& \omega_0(r^2 + a^2) - a p_\varphi. 
\end{eqnarray}
These first-order equations determine the movement of the photons in the presence of plasma. 

\subsection{Spherical photon orbits}

The orbits of photons with constant $r$ should fulfill the two conditions given by $R(r) = R'(r) = 0$; such solutions are unstable, satisfying $R''(r) > 0$. From the equation $R(r) = 0$ we find that
\begin{equation}
	\mathcal{Q} + (p_\varphi-a \omega_0)^2 + f_r = \frac{(\omega_0(r^2+a^2) - a p_\varphi)^2}{\Delta}.
\end{equation}
Then, substituting into $R'(r) = 0$, we obtain
\begin{equation}
	R' = 4\omega_0r(\omega_0(r^2+a^2) - a p_\varphi) -\frac{\Delta'}{\Delta} (\omega_0(r^2+a^2) - a p_\varphi)^2 - \Delta f_r' = 0,
\end{equation}
which is a quadratic equation for $p_\varphi$, with the solution
\begin{equation}
	p_\varphi = \frac{\omega_0}{a}\left[ r^2 + a^2 - \frac{2r\Delta}{\Delta'} \left(1 \pm \sqrt{1 - \frac{\Delta' f_r'}{4 \omega_0^2 r^2}} \right)\right].
\end{equation}
Then, after some algebra, we can obtain $\mathcal{Q}$ in terms of $r$, which reads
\begin{equation}
	\mathcal{Q} = - \frac{\omega_0^2 r^4}{a^2} + \frac{4\omega_0^2 r^2 \Delta}{a^2\Delta'} \left[r - \frac{2}{\Delta'}\left(\Delta - a^2\right)\right] \left(1 \pm \sqrt{1 - \frac{\Delta' f_r'}{4\omega_0^2 r^2}} \right) + \frac{\Delta f_r'}{\Delta' a^2} \left(\Delta - a^2\right) - f_r.
\end{equation}
We have found the critical values of $p_\varphi$ and $\mathcal{Q}$ associated with the spherical orbits of photons. Any trajectory should have $\Theta \geq 0$, then
\begin{equation}
	\mathcal{Q} + \cos^2\theta \left( a^2 \omega_0^2 - \frac{p_\varphi^2}{\sin^2\theta}\right) - f_\theta(\theta) \geq 0.
\end{equation}
This equation defines the photon region: spherical orbits exist at values of $r$ and $\theta$ for which this inequality is satisfied. 

\subsection{Shadow}

For a distant observer, the shadow is determined by the set of photon directions that never reach infinity and instead cross the event horizon. The boundary consists of those rays that asymptotically approach the spherical photon orbits of the spacetime, so they have the same conserved quantities. In order to relate the directions in the sky with the constants of motion, we take an observer at rest in the asymptotically flat region (large $r_\text{o}$) with an inclinaton angle $\theta_\text{o}$ from the spin axis of the black hole, and we construct the orthonormal tetrad as usual\cite{tsukamoto18}:
\begin{equation}
	\mathbf{e}_{\hat{t}} = \partial_t, \qquad
	\mathbf{e}_{\hat{r}} = \partial_r, \qquad
	\mathbf{e}_{\hat{\theta}} = \frac{1}{r_\text{o}} \partial_\theta, \qquad
	\mathbf{e}_{\hat{\varphi}} = \frac{1}{r_\text{o} \sin\theta_\text{o}} \partial_\varphi,
\end{equation}
so that the components of the four-momentum in this frame read
\begin{equation}
	p^{\hat{t}} = \omega_0, \qquad
	p^{\hat{r}} = p^r, \qquad
	p^{\hat{\theta}} = r_\text{o} p^\theta, \qquad
	p^{\hat{\varphi}} = r_\text{o} \sin\theta_\text{o} p^\varphi = \frac{p_\varphi}{r_\text{o} \sin\theta_\text{o}}.
\end{equation}
For the plasma model, we assume that 
\begin{equation}
	\lim_{r \to \infty} \omega_p(r, \theta) =0, 
\end{equation}
which is equivalent to
\begin{equation}
	\lim_{r \to \infty} \frac{f_r(r)}{r^2}=0,
\end{equation}
so photons propagate in vacuum far away from the black hole. We adopt the celestial coordinates for an observer at infinity \cite{bardeen73}
\begin{equation}
	\alpha = - r_\text{o} \frac{p^{\hat{\varphi}}}{p^{\hat{t}}} \bigg|_{r_\text{o} \to \infty}, 
\end{equation}
\begin{equation}
	\beta = - r_\text{o} \frac{p^{\hat{\theta}}}{p^{\hat{t}}} \bigg|_{r_\text{o} \to \infty};
\end{equation}
replacing $p^\theta$ in terms of the conserved quantities, they read
\begin{equation}
	\alpha = - \frac{p_\varphi}{\omega_0 \sin\theta_\text{o}},
\label{alpha} 
\end{equation}
\begin{equation}
	\beta = \pm \frac{1}{\omega_0} \sqrt{\mathcal{Q} + \cos^2\theta_\text{o} \left(a^2 \omega_0^2 - \frac{p_\varphi^2}{\sin^2\theta_\text{o}}\right) - f_\theta(\theta_\text{o})}.
\label{beta}  
\end{equation}
The contour of the black hole shadow is described by the parametric curve $(\alpha(r), \beta(r))$.

\subsection{Observables}

We define three observables \cite{badia20}: the area of the shadow, its oblateness, and the horizontal displacement of its centroid, as follows:
\begin{itemize}
\item The area is calculated from
\begin{equation}
	A = 2 \int \beta\, d\alpha = 2 \int_{r_+}^{r_-} \beta(r) |\alpha'(r)|\, dr,
\end{equation}
with the factor 2 arising from the up-down symmetry of the shadow and taking the plus sign in Eq. (\ref{beta}).
\item The oblateness is related to the deformation of the shadow as compared to a circle
\begin{equation}
	D = \frac{\Delta \alpha}{\Delta \beta},
\end{equation}
where $\Delta\alpha$ and $\Delta\beta$ are the horizontal and vertical extent of the shadow.
\item The horizontal coordinate of the centroid is given by
\begin{equation}
	\alpha_c = \frac{2}{A} \int \alpha \beta\, d\alpha = \frac{2}{A} \int_{r_+}^{r_-} \alpha(r) \beta(r) |\alpha'(r)|\, dr,
\end{equation}
with the same factor 2 as in the definition of the area.
\end{itemize}
These observables allow a characterization of the shadow by only three numbers.

\section{Example: Shapiro plasma}

We consider, in order  to provide an example, the well known model of plasma consisting of dust that is at rest at infinity, introduced by Shapiro \cite{shapiro74}. In the Kerr spacetime, the mass density and the squared plasma frequency go as $r^{-3/2}$, and they are independent of $\theta$ to a very good approximation. However, for this plasma distribution the equations cannot be brought into a separable form; therefore, we take the frequency to have an additional $\theta$ dependency \cite{perlick17} by adopting 
$f_r(r) = \omega_c^2 \sqrt{M^3 r}$ and $f_\theta(\theta) = 0$ , resulting in a plasma electron frequency
\begin{equation}
	\omega_p^2 = \omega_c^2 \frac{\sqrt{M^3 r}}{r^2 + a^2\cos^2\theta},
\end{equation}
where $\omega_c$ is a constant that characterizes the fluid. At large distances from the black hole, the metric approaches the Kerr one, i.e. $m(r) \to M$, so we expect that the Shapiro model is still valid; but at small distances the metric may be quite different, so we assume that the plasma density is not significantly affected.
\begin{figure}[t!]
\includegraphics[width=\textwidth]{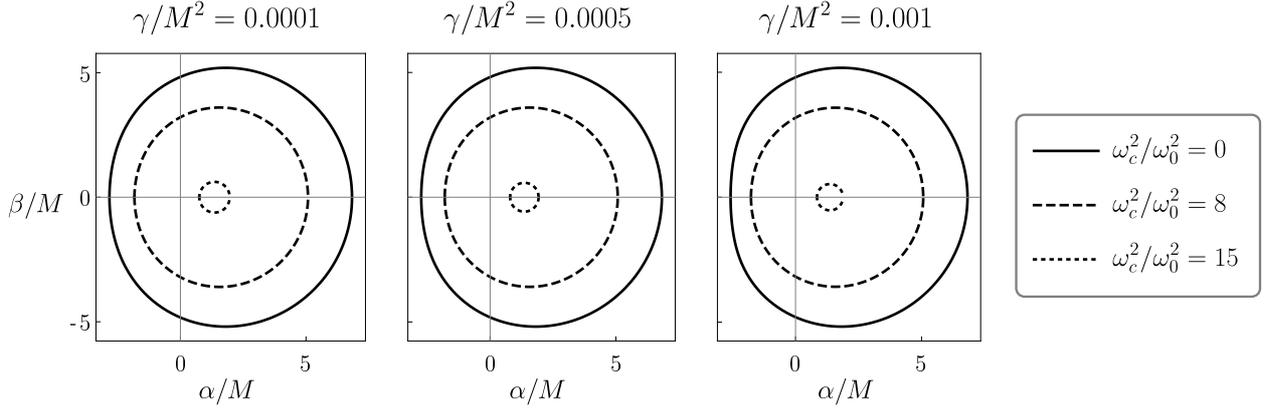}
\caption{Shadow of a $4D$ Einstein-Gauss-Bonnet black hole with spin $a/M=0.9$ surrounded by a Shapiro-type plasma distribution with $f_r(r) = \omega_c^2 \sqrt{M^3 r}$, for an equatorial observer. }
\label{aba:fig1}
\end{figure}
\begin{figure}[t!]
\includegraphics[width=\textwidth]{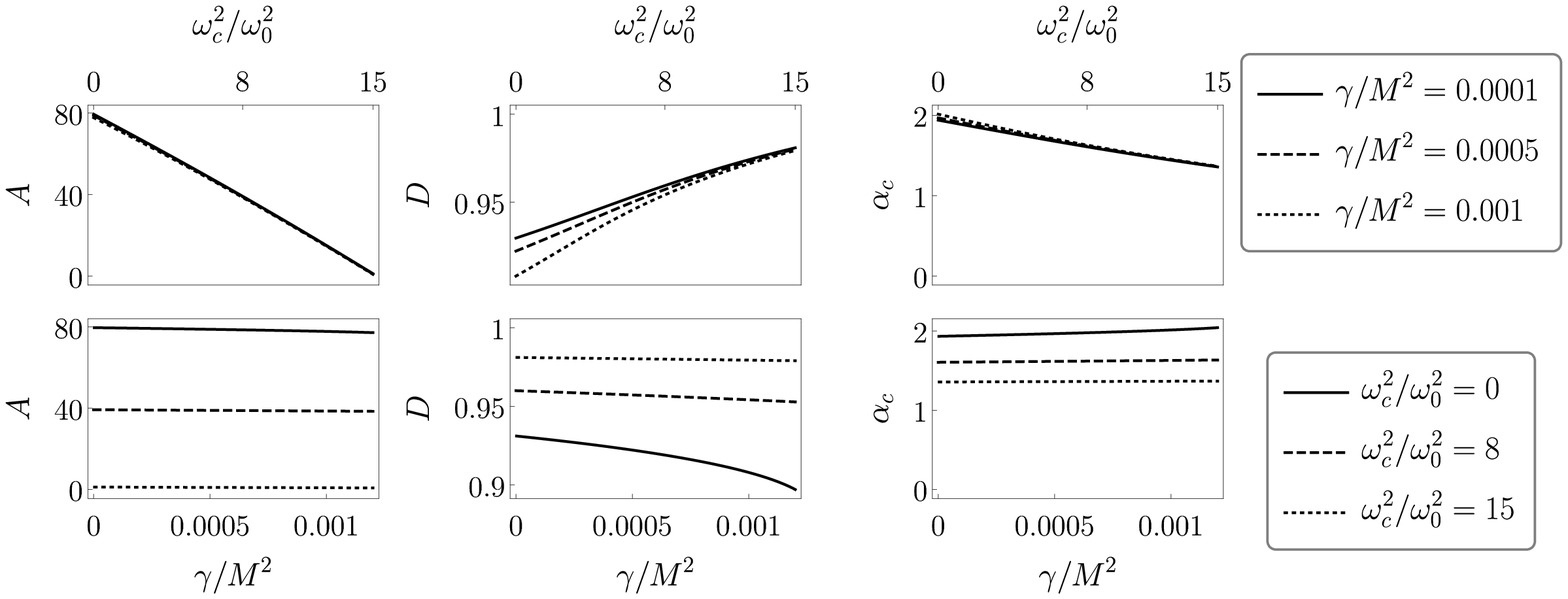}
\caption{The area ($A$), the oblateness ($D$), and the centroid ($\alpha_c$) of the shadow. Top: the observables as functions of the frequency ratio  $\omega_c/\omega_0$, for three values of the EGB parameter $\gamma M^2$. Bottom: the observables as functions of $\gamma M^2$, for three values of the frequency ratio $\omega_c/\omega_0$.}
\label{aba:fig2}
\end{figure}
The trajectories of photons with frequency $\omega_0$ depend only on the ratio $\omega_c/\omega_0$. We present in Figs. \ref{aba:fig1} and \ref{aba:fig2} some plots of  black hole shadows and the corresponding observables, for an equatorial observer, $a/M=0.9$, and suitably chosen values of the other parameters. For a given value of the plasma frequency $\omega_c$, we can see that:
\begin{itemize}
\item For fixed parameter $\gamma/M^2$, the shadow reduces in size, is less deformed, and has a smaller centroid displacement, as the photon frequency $\omega_0$ decreases.
In particular, the shadow disappears entirely below a certain photon frequency, due to the appearance of the forbidden region. This forbidden region starts as two caps around the poles and grows towards the equatorial plane as the frequency decreases, leading to a reduction of the shadow size.
\item For a fixed photon frequency $\omega_0$, the shadow slightly reduces in size, is less deformed, and the centroid displacement slightly grows, as $\gamma/M^2$ increases. The effect of increasing the parameter $\gamma/M^2$ is more prominent for high frequency photons, particularly in the case of the oblateness.
\end{itemize}
The existence of plasma surrounding the black hole results in a smaller and less deformed shadow than in vacuum, which corresponds to $\omega_c =0$. 

\section{Final remarks}

We have studied how the presence of plasma modifies the shadow corresponding to rotating black holes in the scalar-tensor 4-dimensional Einstein-Gauss-Bonnet theory with respect to the vacuum case. These black holes were obtained in previous works from spherically symmetric solutions by applying the Newman-Janis algorithm. We have neglected the gravitational influence of the plasma itself and any processes of scattering, emission or absorption. We have shown that the Hamilton-Jacobi equation for photons is separable if the plasma frequency obeys the usual separability condition introduced for the Kerr spacetime. Light follows timelike curves in the presence of plasma, resulting in a modification of the photon regions and frequency-dependent forbidden regions, where photons cannot travel. We have obtained the expressions for the celestial coordinates of the shadow contour as viewed by a far away observer, which reduce to the already known for vacuum when the plasma frequency goes to zero or the photon frequency to infinity. In order to characterize the size, the shape, and the position, we have presented three observables: the area, the oblateness, and the centroid of the shadow. With the intention to provide a concrete example, we have taken a variation of the Shapiro plasma distribution, which models dust  at rest at infinity surrounding a black hole.  We have considered an equatorial observer, for simplicity and also because the effects on the shadow in this case are more prominent. For a fixed the plasma frequency $\omega_c\neq 0$ and given values of the mass $M$ and the parameter $\gamma$, the shadow becomes smaller and less deformed as the photon frequency $\omega_0$ decreases, with the appearance of a forbidden region around the black hole as a distinctive feature. This forbidden region begins as two caps around the poles and grows towards the equatorial plane for decreasing $\omega_0$, resulting in a sharp reduction of the shadow size and in its eventual disappearance. The presence of plasma always leads to a smaller and less deformed shadow than in the case of vacuum ($\omega_c =0$). The work presented here is a particular case of the shadow cast by a class of rotating black holes surrounded by plasma, which results from applying the Newman-Janis algorithm to spherically symmetric spacetimes \cite{badia21} determined by a mass function $m(r)$ satisfying that $m(r) \to M$ when $r \to \infty$.

\section*{Acknowledgments}

This work has been supported by CONICET and Universidad de Buenos Aires.


\begin{thebibliography}{99}

\bibitem{glavan20} D. Glavan and C. Lin, Einstein-Gauss-Bonnet Gravity in Four-Dimensional Spacetime, {\em Phys. Rev. Lett.} {\bf 124}, 081301 (2020).

\bibitem{gst20} M. Gurses, T.~C. Sisman, and B. Tekin,  Is there a novel Einstein–Gauss–Bonnet theory in four dimensions?, {\em Eur. Phys. J. C} \textbf{80}, 647 (2020).

\bibitem{fernandes20} P.~G.~S. Fernandes, P. Carrilho, T. Clifton, and D.~J. Mulryne, Derivation of regularized field equations for the Einstein-Gauss-Bonnet theory in four dimensions, {\em Phys. Rev. D} {\bf 102}, 024025 (2020).

\bibitem{hennigar20} R.~A. Hennigar, D .Kubiz\v{n}ák, R.~B. Mann, and C. Pollack, On taking the $D \rightarrow 4$ limit of Gauss-Bonnet gravity: theory and solutions, {\em JHEP} {\bf 07} (2020) 027.

\bibitem{bardeen73} J.~M. Bardeen, in {\em Black Holes}, Eds. C. De Witt, B.~S. De Witt, p. 215-239 (1973).

\bibitem{falcke00} H. Falcke, F. Melia, and E. Agol, Viewing the Shadow of the Black Hole at the Galactic Center, {\em Astrophys. J.} {\bf 528}, L13 (2000).

\bibitem{eht19L1} K. Akiyama {\it et al.} (Event Horizon Telescope), First M87 Event Horizon Telescope Results. I. The Shadow of the Supermassive Black Hole, {\em Astrophys. J. Lett.} {\bf 875}, L1 (2019). 

\bibitem{eht19L5} K. Akiyama {\it et al.} (Event Horizon Telescope), First M87 Event Horizon Telescope Results. V. Physical Origin of the Asymmetric Ring, {\em Astrophys. J. Lett.} {\bf 875}, L5 (2019). 

\bibitem{tsukamoto18} N. Tsukamoto, Black hole shadow in an asymptotically flat, stationary, and axisymmetric spacetime: The Kerr-Newman and rotating regular black holes,  {\em Phys. Rev. D} {\bf 97}, 064021 (2018).

\bibitem{badia20}  J. Badía and E.~F. Eiroa, Influence of an anisotropic matter field on the shadow of a rotating black hole, {\em Phys. Rev. D} {\bf 102}, 024066 (2020).

\bibitem{lima20}  H.~C.~D. Lima Junior, L.~C.~B. Crispino, P.~V.~P. Cunha, and C.~A.~R. Herdeiro, Spinning black holes with a separable Hamilton–Jacobi equation from a modified Newman–Janis algorithm, {\em Eur. Phys. J. C} {\bf 80}, 1036 (2020).

\bibitem{perlick21} V. Perlick and O.~Y. Tsupko, Calculating black hole shadows: review of analytical studies, arXiv:2105.07101 [gr-qc].

\bibitem{perlick17} V. Perlick and O.~Y. Tsupko, Light propagation in a plasma on Kerr spacetime: Separation of the Hamilton-Jacobi equation and calculation of the shadow, {\em Phys. Rev. D} {\bf 95}, 104003 (2017).

\bibitem{yan19} H. Yan, Influence of a plasma on the observational signature of a high-spin Kerr black hole, {\em Phys. Rev. D} {\bf 99}, 084050 (2019).

\bibitem{badia21} J. Badía and E.~F. Eiroa, Shadow of axisymmetric, stationary and asymptotically flat black holes in the presence of plasma, {\em Phys. Rev. D} {\bf 104}, 084055 (2021).

\bibitem{kumar20}  R. Kumar and S.~G. Ghosh, Rotating black holes in 4D Einstein-Gauss-Bonnet gravity and its shadow, {\em JCAP} {\bf 07} (2020) 053.

\bibitem{wei21} S. Wei and Y. Liu, Testing the nature of Gauss–Bonnet gravity by four-dimensional rotating black hole shadow, {\em Eur. Phys. J. Plus} {\bf 136}, 436 (2021).

\bibitem{synge60} J.~L. Synge, {\em Relativity: The General Theory} (North-Holland, Amsterdam, 1960).

\bibitem{carter68} B. Carter, Global Structure of the Kerr Family of Gravitational Fields, {\em Phys. Rev.} {\bf 174}, 1559 (1968).

\bibitem{shapiro74} S. Shapiro, Accretion onto black holes: The emergent radiation spectrum. III. Rotating (Kerr) black holes, {\em Astrophys. J.} {\bf 189}, 343 (1974).

\end{thebibliography}
\end{document}